\documentclass[preprint,prd,tightenlines,superscriptaddress]{revtex4-1}

\usepackage{graphicx} 
\usepackage{dcolumn}  
\usepackage{colordvi}
\usepackage{color}
\usepackage{epstopdf}
\usepackage{amssymb}
\usepackage{url}
\graphicspath{{ps}}
\usepackage{hyperref}
\usepackage{tabularx}
\usepackage[normalem]{ulem}
\renewcommand{\arraystretch}{1.1}

\input ./belle2sym.tex

\begin{document}

\vspace*{-3\baselineskip}
\resizebox{!}{3cm}{\includegraphics{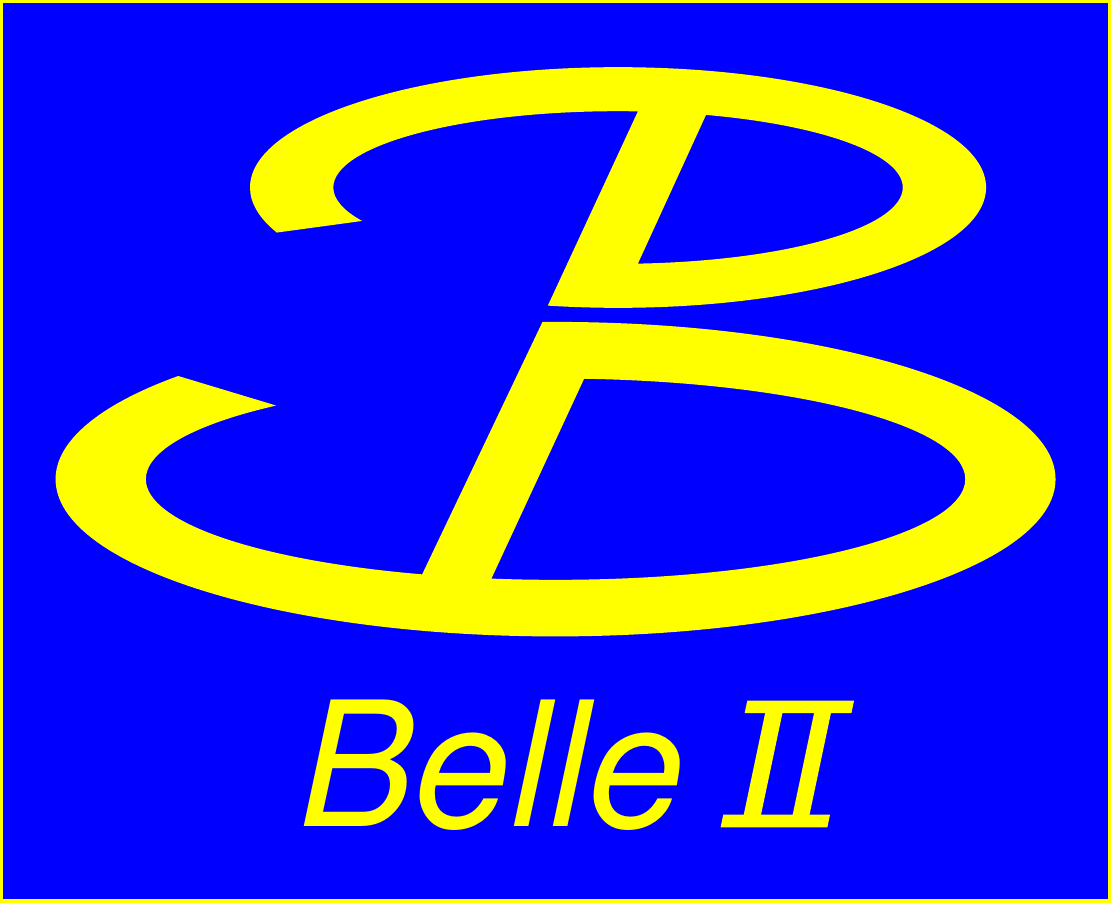}}
\vspace*{-5\baselineskip}

\begin{flushright}
BELLE2-CONF-PH-2021-008 \\
\today
\end{flushright}

\title { \quad\\[0.5cm]  Study of $B\to D^{(*)}h$ decays using $62.8~\mathrm{fb}^{-1}$ of Belle II data}

\newcommand{\instCPPM}{Aix Marseille Universit\'{e}, CNRS/IN2P3, CPPM, 13288 Marseille, France}
\newcommand{\instBeihang}{Beihang University, Beijing 100191, China}
\newcommand{\instBNL}{Brookhaven National Laboratory, Upton, New York 11973, U.S.A.}
\newcommand{\instBINP}{Budker Institute of Nuclear Physics SB RAS, Novosibirsk 630090, Russian Federation}
\newcommand{\instCMU}{Carnegie Mellon University, Pittsburgh, Pennsylvania 15213, U.S.A.}
\newcommand{\instCinvestavIPN}{Centro de Investigacion y de Estudios Avanzados del Instituto Politecnico Nacional, Mexico City 07360, Mexico}
\newcommand{\instPrague}{Faculty of Mathematics and Physics, Charles University, 121 16 Prague, Czech Republic}
\newcommand{\instChiangMai}{Chiang Mai University, Chiang Mai 50202, Thailand}
\newcommand{\instChiba}{Chiba University, Chiba 263-8522, Japan}
\newcommand{\instChonnam}{Chonnam National University, Gwangju 61186, South Korea}
\newcommand{\instConacyt}{Consejo Nacional de Ciencia y Tecnolog\'{\i}a, Mexico City 03940, Mexico}
\newcommand{\instDESY}{Deutsches Elektronen--Synchrotron, 22607 Hamburg, Germany}
\newcommand{\instDuke}{Duke University, Durham, North Carolina 27708, U.S.A.}
\newcommand{\instITAR}{ Duy Tan University, Hanoi 100000, Vietnam}
\newcommand{\instRomaENEA}{ENEA Casaccia, I-00123 Roma, Italy}
\newcommand{\instEri}{Earthquake Research Institute, University of Tokyo, Tokyo 113-0032, Japan}
\newcommand{\instJuelich}{Forschungszentrum J\"{u}lich, 52425 J\"{u}lich, Germany}
\newcommand{\instFuJen}{Department of Physics, Fu Jen Catholic University, Taipei 24205, Taiwan}
\newcommand{\instFudan}{Key Laboratory of Nuclear Physics and Ion-beam Application (MOE) and Institute of Modern Physics, Fudan University, Shanghai 200443, China}
\newcommand{\instGoettingen}{II. Physikalisches Institut, Georg-August-Universit\"{a}t G\"{o}ttingen, 37073 G\"{o}ttingen, Germany}
\newcommand{\instGifu}{Gifu University, Gifu 501-1193, Japan}
\newcommand{\instSOKENDAI}{The Graduate University for Advanced Studies (SOKENDAI), Hayama 240-0193, Japan}
\newcommand{\instGyeongsang}{Gyeongsang National University, Jinju 52828, South Korea}
\newcommand{\instHanyang}{Department of Physics and Institute of Natural Sciences, Hanyang University, Seoul 04763, South Korea}
\newcommand{\instKEK}{High Energy Accelerator Research Organization (KEK), Tsukuba 305-0801, Japan}
\newcommand{\instJPARC}{J-PARC Branch, KEK Theory Center, High Energy Accelerator Research Organization (KEK), Tsukuba 305-0801, Japan}
\newcommand{\instHSE}{Higher School of Economics (HSE), Moscow 101000, Russian Federation}
\newcommand{\instIISER}{Indian Institute of Science Education and Research Mohali, SAS Nagar, 140306, India}
\newcommand{\instIITBhubaneswar}{Indian Institute of Technology Bhubaneswar, Satya Nagar 751007, India}
\newcommand{\instIITGuwahati}{Indian Institute of Technology Guwahati, Assam 781039, India}
\newcommand{\instIITHyderabad}{Indian Institute of Technology Hyderabad, Telangana 502285, India}
\newcommand{\instIITMadras}{Indian Institute of Technology Madras, Chennai 600036, India}
\newcommand{\instIndiana}{Indiana University, Bloomington, Indiana 47408, U.S.A.}
\newcommand{\instIHEPRussia}{Institute for High Energy Physics, Protvino 142281, Russian Federation}
\newcommand{\instHEPHYVienna}{Institute of High Energy Physics, Vienna 1050, Austria}
\newcommand{\instHiroshima}{Hiroshima University, Higashi-Hiroshima, Hiroshima 739-8530, Japan}
\newcommand{\instIHEPChina}{Institute of High Energy Physics, Chinese Academy of Sciences, Beijing 100049, China}
\newcommand{\instIPP}{Institute of Particle Physics (Canada), Victoria, British Columbia V8W 2Y2, Canada}
\newcommand{\instIOP}{Institute of Physics, Vietnam Academy of Science and Technology (VAST), Hanoi, Vietnam}
\newcommand{\instIFIC}{Instituto de Fisica Corpuscular, Paterna 46980, Spain}
\newcommand{\instFrascati}{INFN Laboratori Nazionali di Frascati, I-00044 Frascati, Italy}
\newcommand{\instNapoliINFN}{INFN Sezione di Napoli, I-80126 Napoli, Italy}
\newcommand{\instPadovaINFN}{INFN Sezione di Padova, I-35131 Padova, Italy}
\newcommand{\instPerugiaINFN}{INFN Sezione di Perugia, I-06123 Perugia, Italy}
\newcommand{\instPisaINFN}{INFN Sezione di Pisa, I-56127 Pisa, Italy}
\newcommand{\instRomaINFN}{INFN Sezione di Roma, I-00185 Roma, Italy}
\newcommand{\instRomaTreINFN}{INFN Sezione di Roma Tre, I-00146 Roma, Italy}
\newcommand{\instTorinoINFN}{INFN Sezione di Torino, I-10125 Torino, Italy}
\newcommand{\instTriesteINFN}{INFN Sezione di Trieste, I-34127 Trieste, Italy}
\newcommand{\instJAEA}{Advanced Science Research Center, Japan Atomic Energy Agency, Naka 319-1195, Japan}
\newcommand{\instMainz}{Johannes Gutenberg-Universit\"{a}t Mainz, Institut f\"{u}r Kernphysik, D-55099 Mainz, Germany}
\newcommand{\instGiessen}{Justus-Liebig-Universit\"{a}t Gie\ss{}en, 35392 Gie\ss{}en, Germany}
\newcommand{\instKarlsruhe}{Institut f\"{u}r Experimentelle Teilchenphysik, Karlsruher Institut f\"{u}r Technologie, 76131 Karlsruhe, Germany}
\newcommand{\instISU}{Iowa State University, Ames, Iowa 50011, U.S.A.}
\newcommand{\instKitasato}{Kitasato University, Sagamihara 252-0373, Japan}
\newcommand{\instKISTI}{Korea Institute of Science and Technology Information, Daejeon 34141, South Korea}
\newcommand{\instKoreaUnivKU}{Korea University, Seoul 02841, South Korea}
\newcommand{\instKSU}{Kyoto Sangyo University, Kyoto 603-8555, Japan}
\newcommand{\instKyungpook}{Kyungpook National University, Daegu 41566, South Korea}
\newcommand{\instLPI}{P.N. Lebedev Physical Institute of the Russian Academy of Sciences, Moscow 119991, Russian Federation}
\newcommand{\instLNNU}{Liaoning Normal University, Dalian 116029, China}
\newcommand{\instLMU}{Ludwig Maximilians University, 80539 Munich, Germany}
\newcommand{\instLuther}{Luther College, Decorah, Iowa 52101, U.S.A.}
\newcommand{\instMNITJaipur}{Malaviya National Institute of Technology Jaipur, Jaipur 302017, India}
\newcommand{\instMPP}{Max-Planck-Institut f\"{u}r Physik, 80805 M\"{u}nchen, Germany}
\newcommand{\instMPGHLL}{Semiconductor Laboratory of the Max Planck Society, 81739 M\"{u}nchen, Germany}
\newcommand{\instMcGill}{McGill University, Montr\'{e}al, Qu\'{e}bec, H3A 2T8, Canada}
\newcommand{\instMEPhI}{Moscow Physical Engineering Institute, Moscow 115409, Russian Federation}
\newcommand{\instNagoya}{Graduate School of Science, Nagoya University, Nagoya 464-8602, Japan}
\newcommand{\instNagoyaIAR}{Institute for Advanced Research, Nagoya University, Nagoya 464-8602, Japan}
\newcommand{\instNagoyaKMI}{Kobayashi-Maskawa Institute, Nagoya University, Nagoya 464-8602, Japan}
\newcommand{\instNaraWu}{Nara Women's University, Nara 630-8506, Japan}
\newcommand{\instNTUTaiwan}{Department of Physics, National Taiwan University, Taipei 10617, Taiwan}
\newcommand{\instNUUTaiwan}{National United University, Miao Li 36003, Taiwan}
\newcommand{\instKrakow}{H. Niewodniczanski Institute of Nuclear Physics, Krakow 31-342, Poland}
\newcommand{\instNiigata}{Niigata University, Niigata 950-2181, Japan}
\newcommand{\instNSU}{Novosibirsk State University, Novosibirsk 630090, Russian Federation}
\newcommand{\instOkinawa}{Okinawa Institute of Science and Technology, Okinawa 904-0495, Japan}
\newcommand{\instOsakaCity}{Osaka City University, Osaka 558-8585, Japan}
\newcommand{\instRCNP}{Research Center for Nuclear Physics, Osaka University, Osaka 567-0047, Japan}
\newcommand{\instPNNL}{Pacific Northwest National Laboratory, Richland, Washington 99352, U.S.A.}
\newcommand{\instPanjab}{Panjab University, Chandigarh 160014, India}
\newcommand{\instPanjabPAU}{Punjab Agricultural University, Ludhiana 141004, India}
\newcommand{\instRIKENMSL}{Meson Science Laboratory, Cluster for Pioneering Research, RIKEN, Saitama 351-0198, Japan}
\newcommand{\instSeoul}{Seoul National University, Seoul 08826, South Korea}
\newcommand{\instSPU}{Showa Pharmaceutical University, Tokyo 194-8543, Japan}
\newcommand{\instSoochow}{Soochow University, Suzhou 215006, China}
\newcommand{\instSoongsil}{Soongsil University, Seoul 06978, South Korea}
\newcommand{\instLjubljanaJSI}{J. Stefan Institute, 1000 Ljubljana, Slovenia}
\newcommand{\instKyiv}{Taras Shevchenko National Univ. of Kiev, Kiev, Ukraine}
\newcommand{\instTata}{Tata Institute of Fundamental Research, Mumbai 400005, India}
\newcommand{\instTUM}{Department of Physics, Technische Universit\"{a}t M\"{u}nchen, 85748 Garching, Germany}
\newcommand{\instTelAviv}{Tel Aviv University, School of Physics and Astronomy, Tel Aviv, 69978, Israel}
\newcommand{\instToho}{Toho University, Funabashi 274-8510, Japan}
\newcommand{\instTohoku}{Department of Physics, Tohoku University, Sendai 980-8578, Japan}
\newcommand{\instTitech}{Tokyo Institute of Technology, Tokyo 152-8550, Japan}
\newcommand{\instTokyoMetropolitan}{Tokyo Metropolitan University, Tokyo 192-0397, Japan}
\newcommand{\instUAS}{Universidad Autonoma de Sinaloa, Sinaloa 80000, Mexico}
\newcommand{\instNapoliUNIV}{Dipartimento di Scienze Fisiche, Universit\`{a} di Napoli Federico II, I-80126 Napoli, Italy}
\newcommand{\instPadovaUNIV}{Dipartimento di Fisica e Astronomia, Universit\`{a} di Padova, I-35131 Padova, Italy}
\newcommand{\instPerugiaUNIV}{Dipartimento di Fisica, Universit\`{a} di Perugia, I-06123 Perugia, Italy}
\newcommand{\instPisaUNIV}{Dipartimento di Fisica, Universit\`{a} di Pisa, I-56127 Pisa, Italy}
\newcommand{\instRomaTreUNIV}{Dipartimento di Matematica e Fisica, Universit\`{a} di Roma Tre, I-00146 Roma, Italy}
\newcommand{\instTorinoUNIV}{Dipartimento di Fisica, Universit\`{a} di Torino, I-10125 Torino, Italy}
\newcommand{\instTriesteUNIV}{Dipartimento di Fisica, Universit\`{a} di Trieste, I-34127 Trieste, Italy}
\newcommand{\instMontreal}{Universit\'{e} de Montr\'{e}al, Physique des Particules, Montr\'{e}al, Qu\'{e}bec, H3C 3J7, Canada}
\newcommand{\instIJCLab}{Universit\'{e} Paris-Saclay, CNRS/IN2P3, IJCLab, 91405 Orsay, France}
\newcommand{\instIPHC}{Universit\'{e} de Strasbourg, CNRS, IPHC, UMR 7178, 67037 Strasbourg, France}
\newcommand{\instAdelaide}{Department of Physics, University of Adelaide, Adelaide, South Australia 5005, Australia}
\newcommand{\instBonn}{University of Bonn, 53115 Bonn, Germany}
\newcommand{\instUBC}{University of British Columbia, Vancouver, British Columbia, V6T 1Z1, Canada}
\newcommand{\instCincinnati}{University of Cincinnati, Cincinnati, Ohio 45221, U.S.A.}
\newcommand{\instFlorida}{University of Florida, Gainesville, Florida 32611, U.S.A.}
\newcommand{\instHawaii}{University of Hawaii, Honolulu, Hawaii 96822, U.S.A.}
\newcommand{\instHeidelberg}{University of Heidelberg, 68131 Mannheim, Germany}
\newcommand{\instLjubljanaUniLJ}{Faculty of Mathematics and Physics, University of Ljubljana, 1000 Ljubljana, Slovenia}
\newcommand{\instLouisville}{University of Louisville, Louisville, Kentucky 40292, U.S.A.}
\newcommand{\instMalaya}{National Centre for Particle Physics, University Malaya, 50603 Kuala Lumpur, Malaysia}
\newcommand{\instLjubljanaUM}{University of Maribor, 2000 Maribor, Slovenia}
\newcommand{\instMelbourne}{School of Physics, University of Melbourne, Victoria 3010, Australia}
\newcommand{\instMississippi}{University of Mississippi, University, Mississippi 38677, U.S.A.}
\newcommand{\instUOM}{University of Miyazaki, Miyazaki 889-2192, Japan}
\newcommand{\instPittsburgh}{University of Pittsburgh, Pittsburgh, Pennsylvania 15260, U.S.A.}
\newcommand{\instUSTC}{University of Science and Technology of China, Hefei 230026, China}
\newcommand{\instSAlabama}{University of South Alabama, Mobile, Alabama 36688, U.S.A.}
\newcommand{\instSCarolina}{University of South Carolina, Columbia, South Carolina 29208, U.S.A.}
\newcommand{\instSydney}{School of Physics, University of Sydney, New South Wales 2006, Australia}
\newcommand{\instUTokyo}{Department of Physics, University of Tokyo, Tokyo 113-0033, Japan}
\newcommand{\instIPMU}{Kavli Institute for the Physics and Mathematics of the Universe (WPI), University of Tokyo, Kashiwa 277-8583, Japan}
\newcommand{\instVictoria}{University of Victoria, Victoria, British Columbia, V8W 3P6, Canada}
\newcommand{\instVPI}{Virginia Polytechnic Institute and State University, Blacksburg, Virginia 24061, U.S.A.}
\newcommand{\instWayneState}{Wayne State University, Detroit, Michigan 48202, U.S.A.}
\newcommand{\instYamagata}{Yamagata University, Yamagata 990-8560, Japan}
\newcommand{\instYerevan}{Alikhanyan National Science Laboratory, Yerevan 0036, Armenia}
\newcommand{\instYonsei}{Yonsei University, Seoul 03722, South Korea}
\newcommand{\instZZU}{Zhengzhou University, Zhengzhou 450001, China}
\affiliation{\instCPPM}
\affiliation{\instBeihang}
\affiliation{\instBNL}
\affiliation{\instBINP}
\affiliation{\instCMU}
\affiliation{\instCinvestavIPN}
\affiliation{\instPrague}
\affiliation{\instChiangMai}
\affiliation{\instChiba}
\affiliation{\instChonnam}
\affiliation{\instConacyt}
\affiliation{\instDESY}
\affiliation{\instDuke}
\affiliation{\instITAR}
\affiliation{\instRomaENEA}
\affiliation{\instEri}
\affiliation{\instJuelich}
\affiliation{\instFuJen}
\affiliation{\instFudan}
\affiliation{\instGoettingen}
\affiliation{\instGifu}
\affiliation{\instSOKENDAI}
\affiliation{\instGyeongsang}
\affiliation{\instHanyang}
\affiliation{\instKEK}
\affiliation{\instJPARC}
\affiliation{\instHSE}
\affiliation{\instIISER}
\affiliation{\instIITBhubaneswar}
\affiliation{\instIITGuwahati}
\affiliation{\instIITHyderabad}
\affiliation{\instIITMadras}
\affiliation{\instIndiana}
\affiliation{\instIHEPRussia}
\affiliation{\instHEPHYVienna}
\affiliation{\instHiroshima}
\affiliation{\instIHEPChina}
\affiliation{\instIPP}
\affiliation{\instIOP}
\affiliation{\instIFIC}
\affiliation{\instFrascati}
\affiliation{\instNapoliINFN}
\affiliation{\instPadovaINFN}
\affiliation{\instPerugiaINFN}
\affiliation{\instPisaINFN}
\affiliation{\instRomaINFN}
\affiliation{\instRomaTreINFN}
\affiliation{\instTorinoINFN}
\affiliation{\instTriesteINFN}
\affiliation{\instJAEA}
\affiliation{\instMainz}
\affiliation{\instGiessen}
\affiliation{\instKarlsruhe}
\affiliation{\instISU}
\affiliation{\instKitasato}
\affiliation{\instKISTI}
\affiliation{\instKoreaUnivKU}
\affiliation{\instKSU}
\affiliation{\instKyungpook}
\affiliation{\instLPI}
\affiliation{\instLNNU}
\affiliation{\instLMU}
\affiliation{\instLuther}
\affiliation{\instMNITJaipur}
\affiliation{\instMPP}
\affiliation{\instMPGHLL}
\affiliation{\instMcGill}
\affiliation{\instMEPhI}
\affiliation{\instNagoya}
\affiliation{\instNagoyaIAR}
\affiliation{\instNagoyaKMI}
\affiliation{\instNaraWu}
\affiliation{\instNTUTaiwan}
\affiliation{\instNUUTaiwan}
\affiliation{\instKrakow}
\affiliation{\instNiigata}
\affiliation{\instNSU}
\affiliation{\instOkinawa}
\affiliation{\instOsakaCity}
\affiliation{\instRCNP}
\affiliation{\instPNNL}
\affiliation{\instPanjab}
\affiliation{\instPanjabPAU}
\affiliation{\instRIKENMSL}
\affiliation{\instSeoul}
\affiliation{\instSPU}
\affiliation{\instSoochow}
\affiliation{\instSoongsil}
\affiliation{\instLjubljanaJSI}
\affiliation{\instKyiv}
\affiliation{\instTata}
\affiliation{\instTUM}
\affiliation{\instTelAviv}
\affiliation{\instToho}
\affiliation{\instTohoku}
\affiliation{\instTitech}
\affiliation{\instTokyoMetropolitan}
\affiliation{\instUAS}
\affiliation{\instNapoliUNIV}
\affiliation{\instPadovaUNIV}
\affiliation{\instPerugiaUNIV}
\affiliation{\instPisaUNIV}
\affiliation{\instRomaTreUNIV}
\affiliation{\instTorinoUNIV}
\affiliation{\instTriesteUNIV}
\affiliation{\instMontreal}
\affiliation{\instIJCLab}
\affiliation{\instIPHC}
\affiliation{\instAdelaide}
\affiliation{\instBonn}
\affiliation{\instUBC}
\affiliation{\instCincinnati}
\affiliation{\instFlorida}
\affiliation{\instHawaii}
\affiliation{\instHeidelberg}
\affiliation{\instLjubljanaUniLJ}
\affiliation{\instLouisville}
\affiliation{\instMalaya}
\affiliation{\instLjubljanaUM}
\affiliation{\instMelbourne}
\affiliation{\instMississippi}
\affiliation{\instUOM}
\affiliation{\instPittsburgh}
\affiliation{\instUSTC}
\affiliation{\instSAlabama}
\affiliation{\instSCarolina}
\affiliation{\instSydney}
\affiliation{\instUTokyo}
\affiliation{\instIPMU}
\affiliation{\instVictoria}
\affiliation{\instVPI}
\affiliation{\instWayneState}
\affiliation{\instYamagata}
\affiliation{\instYerevan}
\affiliation{\instYonsei}
\affiliation{\instZZU}
  \author{F.~Abudin{\'e}n}\affiliation{\instTriesteINFN} 
  \author{I.~Adachi}\affiliation{\instKEK}\affiliation{\instSOKENDAI} 
  \author{R.~Adak}\affiliation{\instFudan} 
  \author{K.~Adamczyk}\affiliation{\instKrakow} 
  \author{P.~Ahlburg}\affiliation{\instBonn} 
  \author{J.~K.~Ahn}\affiliation{\instKoreaUnivKU} 
  \author{H.~Aihara}\affiliation{\instUTokyo} 
  \author{N.~Akopov}\affiliation{\instYerevan} 
  \author{A.~Aloisio}\affiliation{\instNapoliUNIV}\affiliation{\instNapoliINFN} 
  \author{F.~Ameli}\affiliation{\instRomaINFN} 
% ...
\collaboration{Belle II Collaboration}

\begin{abstract}
We report measurements related to hadronic $B$ decays to final states that contain charm mesons. The analyses are performed on a
$62.8~\mathrm{fb}^{-1}$ data set collected by the Belle II experiment at a center-of-mass energy corresponding to the mass of the $\Upsilon(4S)$ resonance. The measurements reported are for the decay modes $B^-\to D^0 h^-$, $B^{-}\to D^{*0}h^-$, $\bar{B}^{0}\to D^{+} h^{-}$ and $\bar{B}^{0}\to D^{*+} h^{-}$, where $h=\pi$ or $K$. These modes are either signal or control channels for measurements related to the unitarity triangle angle $\gamma$ in direct or time-dependent $CP$-violation measurements. The reported observables are the ratios between the $B\to D^{(*)}K$ and $B\to D^{(*)}\pi$ decay rates, which are found to be in agreement with previous measurements.
\end{abstract}

\maketitle


\section{Introduction}
We report the first measurements at Belle II of observables related to $B^-\to D^{(*)0}h^-$ and $\bar{B}^0\to D^{(*)+}h^-$ decays, where $h^{-}$ is either a $\pi^-$ or $K^{-}$ meson. (Throughout this paper charge-conjugate is implied.) These decay modes are of interest for two reasons. Firstly, the decays $B^-\to D^{(*)0}\pi^-$ and $\bar{B}^0\to D^{(*)+}\pi^-$ arise from the favoured $b\to c$ transition, which makes them some of the most abundant hadronic $B$ decays with branching fractions between 0.25\% and 0.5\%~\cite{pdg}. Therefore, these modes are important control channels for other fully hadronic $B$-decay measurements, such as those of time-dependent $CP$ violation and charmless $B$ decays. Secondly, the decays $B^{-}\to D^{(*)0}K^-$ are sensitive to the $b-d$ Cabibbo-Kobayashi-Maskawa (CKM)~\cite{ckm} unitarity-triangle angle $\phi_3$ (or $\gamma$)~\cite{GLW}. A more precise determination of $\phi_3$ is one of the primary goals of Belle II~\cite{b2tip}. 

An important set of observables related to these modes are the ratios between the decay rates:
\begin{eqnarray}
R^{(*)0} & = & \frac{\Gamma(B^-\to D^{(*)0}K^-)}{\Gamma(B^-\to D^{(*)0}\pi^-)} \\
R^{(*)+} & = & \frac{\Gamma(\bar{B}^0\to D^{(*)+}K^-)}{\Gamma(\bar{B}^0\to D^{(*)+}\pi^-)} \;.
\label{eq:ratio}
\end{eqnarray}
These observables can test theoretical predictions, particularly of factorization and $SU(3)$ symmetry breaking in quantum chromodynamics (QCD)~\cite{RQCD}. We present measurements of $R^{(*)0/+}$ for four decay modes: (1) $B^{-}\to D^0 h^-$, $D^{0}\to K^-\pi^+$ or $D^0\to K^{0}_{\rm S}\pi^+\pi^-$; (2) $B^{-}\to D^{*0}h^-$, $D^{*0}\to D^{0}\pi^0$, $D^{0}\to K^{-}\pi^+$; (3) $\bar{B}^{0}\to D^+ h^-$, $D^+\to K^-\pi^+\pi^+$; and (4) $\bar{B}^{0}\to D^{*+} h^-$, $D^{*+}\to D^0\pi^+$, $D^0\to K^-\pi^+$. 

 Among these decays $B^-\to D^0(K^0_{\rm S}\pi^+\pi^-)K^-$ is the single most sensitive mode to determine $\phi_3$ 
\cite{bondarK0Spipi,giri,bellegamma} at Belle II \cite{b2tip}. Therefore, the demonstration of an efficient reconstruction of this mode at Belle II is a significant first step toward a determination of $\phi_3$. Hence, a more complex analysis is performed for $B^-\to D^0(K^0_{\rm S}\pi^+\pi^-)K^-$ compared to the other modes, which are used as high-statistics control samples or for tests of QCD.

The remainder of this paper is organised as followed. Section~\ref{sec:detector} describes the Belle II detector, as well as the data and simulation samples used in these analyses. The event selection requirements are outlined in Sec. \ref{sec:select}. Section~\ref{sec:results} describes how the values of $R^{(*)0/+}$ are determined from the data, the results are presented and the evaluation of systematic uncertainties described.
Section~\ref{sec:summary} gives the conclusion and outlook.

\section{The Belle~II detector and data sample} \label{sec:detector}

Belle II~\cite{belleII} is a  particle-physics spectrometer with almost $4\pi$ solid-angle coverage, which is designed to reconstruct the products of electron-positron collisions produced by the SuperKEKB asymmetric-energy collider~\cite{superkekb}, located at the KEK laboratory in Tsukuba, Japan. The energies of the electron and positron beams are 7~GeV and 4~GeV, respectively. Belle II comprises several subdetectors arranged around the interaction point in a cylindrical geometry. The innermost subdetector is the vertex detector  (VXD), which uses position-sensitive silicon layers to sample the trajectories of charged particles (tracks) in the vicinity of the interaction region to extrapolate the decay positions of their long-lived parent particles. 
The  VXD includes two inner layers of  DEPFET-based pixel sensors and four outer layers of double-sided silicon microstrip sensors. The second pixel layer is currently incomplete covering only  one sixth of the azimuthal angle. Charged-particle momenta and charges are measured by a large-radius, helium-ethane, small-cell central drift chamber (CDC), which also offers charged-particle-identification information through a measurement of particles'  specific ionization. A Cherenkov-light angle and time-of-propagation (TOP) detector surrounding the chamber provides charged-particle identification in the central detector volume, supplemented by proximity-focusing, aerogel, ring-imaging Cherenkov (ARICH) detectors in the forward  region with respect to the electron beam. 
A CsI(Tl)-crystal electromagnetic calorimeter (ECL)  provides electron-energy measurements and photon reconstruction. 
A solenoid surrounding the calorimeter generates a uniform axial 1.5~T magnetic field filling its inner volume.
Layers of plastic scintillator and resistive-plate chambers, interspersed between the magnetic flux-return iron plates, allow for identification of $K^0_{\rm L}$ and muons. 
The subdetectors most relevant for this work are the VXD, CDC, TOP, ARICH and ECL.

We use simulated data to optimize the event selection,  study background and compare the distributions observed in experimental data with expectations. 
We use signal-only simulated data to model relevant signal features for fits and determine selection efficiencies. 
 The so-called {\it generic} sample consists of Monte Carlo (MC)  simulated events that include $B^0{\bar{B}}^0$, $B^+B^-$, $u\bar{u}$, $d\bar{d}$, $s\bar{s}$, and $c\bar{c}$ processes in realistic proportions and corresponds in size to more than ten times that of the $\Upsilon(4S)$ data.  The generic MC sample is used to study background and make comparisons with the data.
In addition, one million signal-only events are generated for each decay channel. The $B$-meson decays are simulated with the {\sc EvtGen} generator \cite{evtgen} and the effect of final-state electromagnetic radiation is simulated by the {\sc Photos} package \cite{photos}. The simulation of the continuum background process $e^+ e^-\to q\bar{q}~(q = u,~d,~s,~c)$ is carried out with the {\sc KKMC}~\cite{kkmc} generator interfaced to {\sc Pythia}~\cite{pythia}. The interactions of particles with the detector are simulated using {\sc Geant4}~\cite{geant4}.

The data sample consists of all good-quality runs collected at a center-of-mass energy corresponding to the the $\Upsilon(4S)$ resonance  from March 11$^{\rm th}$, 2019 until July~1$^{\rm st}$, 2020. The sample size corresponds to an integrated luminosity of $62.8~\mathrm{fb}^{-1}$. 
Events used in the analysis are required to satisfy data-skimming selection criteria, which reduces sample sizes such that the time required to analyse the full data sample is shortened significantly. These skimming criteria are placed on the total energy and charged-particle multiplicity in the event; the selection is almost 100\% efficient on signal events  rejecting only beam background and low-multiplicity events, such as those produced in two-photon collisions. All data are processed using the Belle II analysis software framework~\cite{basf2}.

\section{Event selection and reconstruction} \label{sec:select}

The selection has been designed to be largely common among the modes studied. An overview of the selection is as follows. Initially we select $\pi^+$, $K^+$, $\pi^0$ and $K^{0}_{\rm S}$ candidates with baseline  criteria that ensure high efficiency and purity. These candidates are combined to form $D$ and $D^*$ candidates, which are then combined with an $h^-$ candidate to form $B$ candidates. Constrained vertex and kinematic fits are applied to ensure consistency with the topology of the decay. We reconstruct $B^-\to D^0 h^-$, $B^{-}\to D^{*0}h^-$, $\bar{B}^{0}\to D^{+} h^{-}$ and $\bar{B}^{0}\to D^{*+} h^{-}$. Further, we reconstruct $D^{*+}\to D^{0} \pi^+$ and  $D^{*0}\to D^{0} \pi^0$ cascades, as well as $D^{0} \to K^{-}\pi^{+}$, $K^0_{\rm S} \pi^{+} \pi^{-}$ and $D^{+} \to K^{-}\pi^{+}\pi^{+}$ decays.  The remainder of this section describes the details of the selection criteria.

Charged particle tracks originating from $e^+e^-$ collisions are selected by requiring $|dr|<0.5$~cm and $|dz|<3$~cm, where $dr$ and $dz$ represent the distance of closest approach to the interaction point (IP) in the plane transverse to and along the  $z$ direction, respectively. (In the Belle II coordinate system, the $z$ axis is aligned with the direction opposite to the positron beam.) These tracks are then identified as either $K^+$ or $\pi^+$ using information from the CDC, TOP and ARICH detectors. We apply likelihood-ratio requirements of $\mathcal{L}(K/\pi) =
\frac{\mathcal{L}_{K}}{\mathcal{L}_{K}+\mathcal{L}_{\pi}} > 0.6$ for a kaon candidate and
$\mathcal{L}(K/\pi) < 0.6$ for a pion candidate, where $\mathcal{L}_{K}$ ($\mathcal{L}_{\pi}$) is the likelihood of a track being a kaon (pion).
In order to reduce the pion fake rate,  we require $\cos\theta > -0.6$, where $\theta$ is the polar angle in lab frame of the $\pi$ or $K$ candidate coming directly from the $B$ decay, which is referred to as the {\it prompt} track. This requirement removes the tracks in the backward part of the detector, which  are outside the TOP or ARICH acceptance.  The kaon identification efficiency is approximately 83\% and the pion-to-kaon misidentifiaction rate is about 10\%; the latter quantity is obtained from the simultaneous fit to $B\to D^{(*)}h^-$ data described in Sec.~\ref{sec:results}.

For $K^0_{\rm S}$ reconstruction, we use pairs of oppositely  charged tracks that originate from a common space-point and have an  invariant mass consistent with the nominal $K^0_{\rm S}$ mass \cite{pdg},  when the tracks are reconstructed assuming the pion mass hypothesis. No particle identification criteria are applied to these tracks. To improve the purity of the $K_{\rm S}^0$ selection, we use a multivariate discriminant implemented as a fast boosted decision tree (FBDT). Five variables are used as inputs to the FBDT:
\begin{enumerate}
\item{The azimuthal angle between the momentum vector and the vector between the IP and the decay vertex of the $K^0_{\rm S}$ candidate.}
\item{The smaller distance of closest approach between the extrapolated track of one of the pion candidates and the IP.}
\item{The longer distance of closest approach between the extrapolated track of the other pion candidate and the IP.}
\item{The flight length of the $K_{\rm S}^0$ candidate in the plane transverse to the beam direction.}
\item{The difference between the measured mass of $K^0_{\rm S}$ candidate and the nominal $K^0_{\rm S}$ mass~\cite{pdg} divided by uncertainty on the measured $K^0_{\rm S}$ candidate mass.} 
\end{enumerate}
 The efficiency and purity of the $K^{0}_{\rm S}$ selection are 91\% and 97\%, respectively.

We reconstruct $\pi^0$ candidates from photon pairs. The energy of each photon is required to be greater than 30, 80, and 60~MeV depending upon whether it is reconstructed in the barrel, forward, and backward endcap region of the ECL, respectively;  the differing thresholds are motivated by the different levels of beam-induced background within the regions. Requirements are placed on the helicity angle of the $\pi^0$ decay to further reduce combinatorial candidates constructed from the beam-induced background. We also restrict the diphoton mass to be between $120 < M_{\gamma\gamma} < 145~\mathrm{MeV}/c^{2}$. The mass of the $\pi^0$ candidates is constrained to its known value in subsequent kinematic fits to improve its four-momentum resolution. 

Invariant-mass restrictions are placed on the the $D$ and $D^*$ candidates, formed from combinations of the selected $\pi^+$, $K^+$, $K^0_{\rm S}$ and $\pi^0$ candidates, to reduce combinatorial background:
  \begin{itemize} 
 \item $1.84~<~M (K^{-}\pi^{+})~<~1.89$ GeV/$c^{2}$;
 \item $1.85~<~M (K_{\rm S}^{0}\pi^{-}\pi^{+})~<~1.88$ GeV/$c^{2}$;
  \item $ 1.844~<~M(K^{-}\pi^{+}\pi^{+})~<~1.894$ GeV/$c^{2}$;
 \item $0.140~<$ $M(D^0\pi^{0}) - M(D^0)~<~0.144$ GeV/$c^{2}$; and 
  \item $0.143~<$ $M(D^0\pi^{+}) - M(D^0)~<~0.147$ GeV/$c^{2}$.
  \end{itemize}
  These intervals correspond to between $\pm 3.5\sigma$ and $\pm 4.0\sigma$ about the nominal $D^{(*)}$ masses \cite{pdg}, where $\sigma$ is the invariant-mass resolution. To improve the resolution of the selected $D^{(*)}$ candidate's four-momentum, it is reconstructed using a kinematic fit that constraints the reconstructed mass to the known $D^{(*)}$ mass \cite{pdg}.  This fit improves the resolution of the beam-energy difference [defined in Eq.~(\ref{eq:DE})] by approximately 11\%. 

$B$-meson candidates are reconstructed by combining a $D$ or $D^*$ candidate  with a charged track without any particle identification criteria applied. The kinematic variables used to discriminate $B$ decays from combinatorial or partially reconstructed background are the beam-energy-constrained mass 
\begin{equation}
M_{\rm bc} = \frac{1}{c^2}\sqrt{E^2_{\rm beam} - \left(\sum_i {\vec{\mathbf{p}}}_ic\right)^2}\;,
\end{equation}
and the beam-energy difference
\begin{equation}\label{eq:DE}
     \Delta E = \sum_i E_i - E_{\rm beam}\;,
\end{equation}
where $E_{\rm beam}$ is the beam energy  and $(E_i,\vec{\mathbf{p}}_ic)$ is the four-momentum of the $i^{\rm th}$ decay product of the $B$ candidate;  all quantities are calculated in the center-of-mass frame. For correctly reconstructed signal, $M_{\rm bc}$ peaks at the nominal mass of the $B$ meson \cite{pdg}
and $\Delta E$ peaks at zero. We  retain candidates with $M_{\rm bc} > 5.27$~GeV/$c^2$. The distribution of $\Delta E$ for selected candidates is fit to determine the values of $R^{(*)+/0}$. Therefore, mode-dependent $\Delta E$ criteria are placed to define the interval over which the distribution is fit. The reason for the differing $\Delta E$ ranges is two-fold: to remove partially reconstructed background and to increase sideband control regions for the two-dimensional fit in the mode $B^- \to D(K_{\rm S}^0\pi^+\pi^-) h^-$. The $\Delta E$ criteria are $-0.13<\Delta E < 0.15$~GeV for $B^-\to D^{0}(K^-\pi^+)h^-$ and $B^-\to D^{*0}[D^0(K^-\pi^+)\pi^0]h^-$, $-0.13<\Delta E < 0.18$~GeV for $B^-\to D^{0}(K^0_{\rm S}\pi^+\pi^-)h^-$, and  $-0.15<\Delta E < 0.15$~GeV for $\bar{B}^0\to D^{(*)+}h^-$.


There are a few background modes that can peak in the  same manner as the signal (‘peaking background’). The decay $B^- \to J/\psi (\ell^+ \ell^-)K^-$ may contribute to the background for $B^- \to D^0(K^-\pi^+)\pi^-$ or similarly, $\bar{B}^0 \to J/\psi (\ell^+ \ell^-)K^{*0}$ for $\bar{B}^0 \to D^{+}(K^-\pi^+\pi^+)\pi^-$. To reject this background  arising from particle misidentification, we veto candidates satisfying $M(\pi\pi)$ being within $\pm3\sigma$ of the nominal $J/\psi$ mass.

Continuum background is suppressed by requiring the ratio of second and zeroth Fox-Wolfram moments~\cite{fox}, $R_2$ $<$ 0.3. This  criterion is applied for all modes except for $B^- \to D (K_{\rm S}^0\pi^+\pi^-) h^-$, for which we use instead  an FBDT that combines variables known to provide statistical discrimination between $B$-meson signals and continuum background. The variables are also required to have negligible correlation with $\Delta E$ and $M_{\rm bc}$.  These quantities are associated to event topology, which relate to both the whole event and signal-only angular configurations. 
We train the classifier to identify statistically significant signal and background features using  simulated samples.
We use the following event topology variables for differentiating the signal and continuum: the likelihood ratio obtained from Fisher discriminants formed from modified Super-Fox-Wolfram moment \cite{skfw}, the absolute value of the cosine of the angle between the $B$ candidate and the $z$ direction in the $e^{+}e^{-}$ center-of-mass frame, the cosine of the angle between the thrust axis of the signal $B$ and thrust axis of rest-of-the-event (ROE), the difference between the position of the signal $B$ decay vertex and the vertex of the ROE in the $z$ direction, and  $B$ flavor-tagger output~\cite{ft}. 
The output of the FBDT lies in the range zero to one, where signal events peak around one and continuum events peak around zero. 
The fit to data in this mode includes a variable related to the FBDT output as well as $\Delta E$. However, to simplify the background description of this variable, the FBDT output is required to be greater than 0.2; this criterion rejects 67\% of background while retaining 96\% of signal.  

After applying all the selection criteria, there can be more than one candidate per event in the $B^- \to D (K_{\rm S}^0\pi^+\pi^-) h^-$ mode with an average multiplicity of approximately 1.06. In such events we retain the candidate that has $M (K_{\rm S}^{0}\pi^{+}\pi^{-})$ and $M_{\rm bc}$ values closest to the corresponding nominal values \cite{pdg};  the efficiency of this criterion is approximately 65\%. The number of events with multiple candidates is negligible in all other decay modes studied. 

\section{Results}
\label{sec:results}
We select both $B \to DK$ and $B \to D\pi$ decays, the $B \to D\pi$ branching fraction is typically an order of magnitude larger than that of $B \to DK$,  hence it can serve as an excellent calibration sample for the signal determination procedure.
Furthermore, there is a significant background from $B \to D \pi$ decays in the $B \to DK$ sample  due to the misidentification of the charged pion as a kaon; a simultaneous fit to samples enhanced in prompt tracks that are either pions $[L(K/\pi)<0.6]$ or kaons $[L(K/\pi)>0.6]$, allows this cross-feed to be directly determined from data.
The signal extraction is done by fitting only the $\Delta E$ distribution simultaneously in pion and kaon enhanced samples for all the modes other than $B^- \to D^0h^-$ where $D^0 \to K_{\rm S}^0\pi^+\pi^-$. 
In this last case, the signal yield is determined from a two-dimensional extended maximum-likelihood fit to $\Delta E$ and the transformed FBDT output ($C'$). 
The continuum suppression FBDT output is transformed using the $\mu$-transformation~\cite{mutrans1, mutrans2}. 
 The principal advantage to using this transformation compared to the commonly used Gaussian transformation (see for example Ch. 9 in Ref.~\cite{bfactory}) is that the PDFs can be described by analytic functions that have fewer parameters.

The yields of the signal $B \to D^{(*)}\pi$ and $B \to D^{(*)}K$, and their cross-feed, in the pion and kaon-enhanced samples can be expressed by the following relations:
\begin{eqnarray}
    N^{D^{(*)}\pi}_{\rm pion \; enhanced} & = & (1-\kappa)\, N^{D^{(*)}\pi}_{\rm tot} \\
    N^{D^{(*)}\pi}_{\rm kaon \; enhanced} & = & \kappa \, N^{D^{(*)}\pi}_{\rm tot} \\
    N^{D^{(*)}K}_{\rm kaon \; enhanced} & = &\epsilon \, R^{(*)} \, N^{D^{(*)}\pi}_{\rm tot} \\
   N^{D^{(*)}K}_{\rm pion \; enhanced} & = & (1- \epsilon) \, R^{(*)} \,N^{D^{(*)}\pi}_{\rm tot}\;.
\end{eqnarray}
Here the pion fake rate $\kappa$ is a free parameter, as well as $R^{(*)}$ and $N^{D^{(*)}\pi}_{\rm tot}$, respectively the ratio between the decay rates defined in Eq.~(\ref{eq:ratio}) and the signal yield of $B\to D^{(*)}\pi$ mode. Due to the low yield of $B \to D^{(*)}K$ cross feed to the pion-enhanced sample, the kaon identification efficiency $\epsilon$ is fixed to the value obtained from the tagged $D$ control samples that are used to calibrate the particle identification~\cite{pid_belle2}  performance. 
\begin{figure}[!tb]
\centering
\begin{tabular}{c c}
    \includegraphics[scale=0.4]{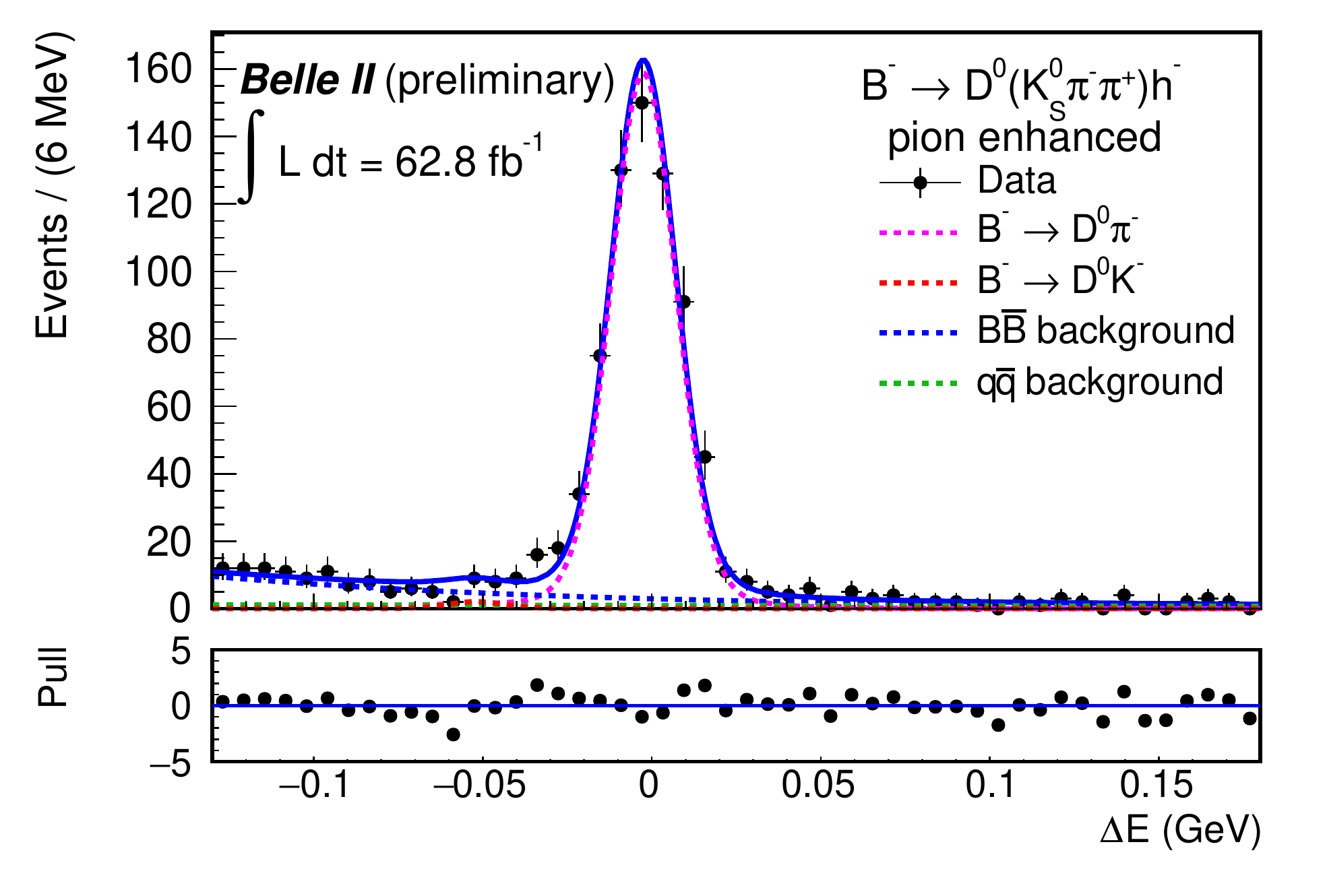} &
    \includegraphics[scale=0.4]{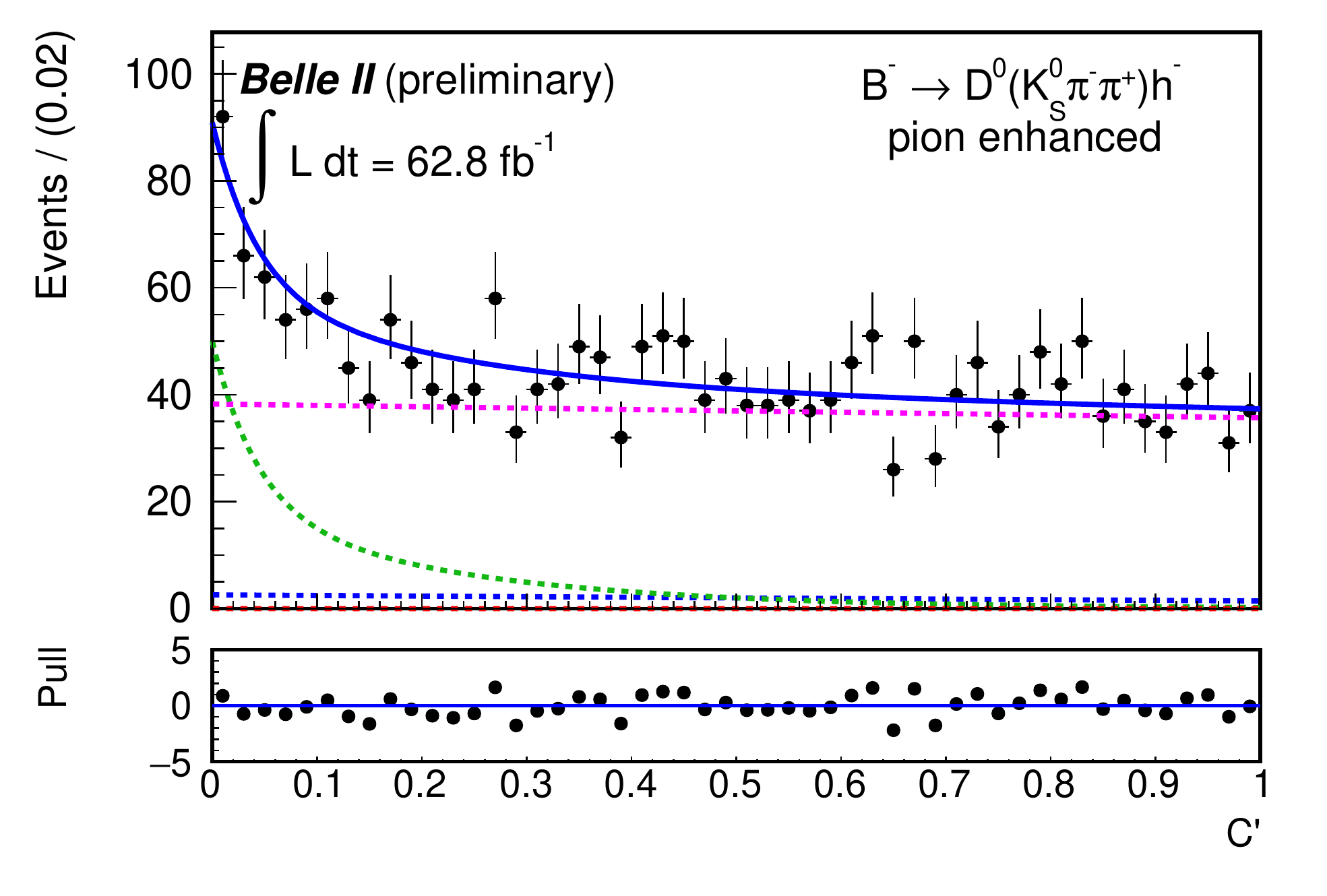} \\
    \includegraphics[scale=0.4]{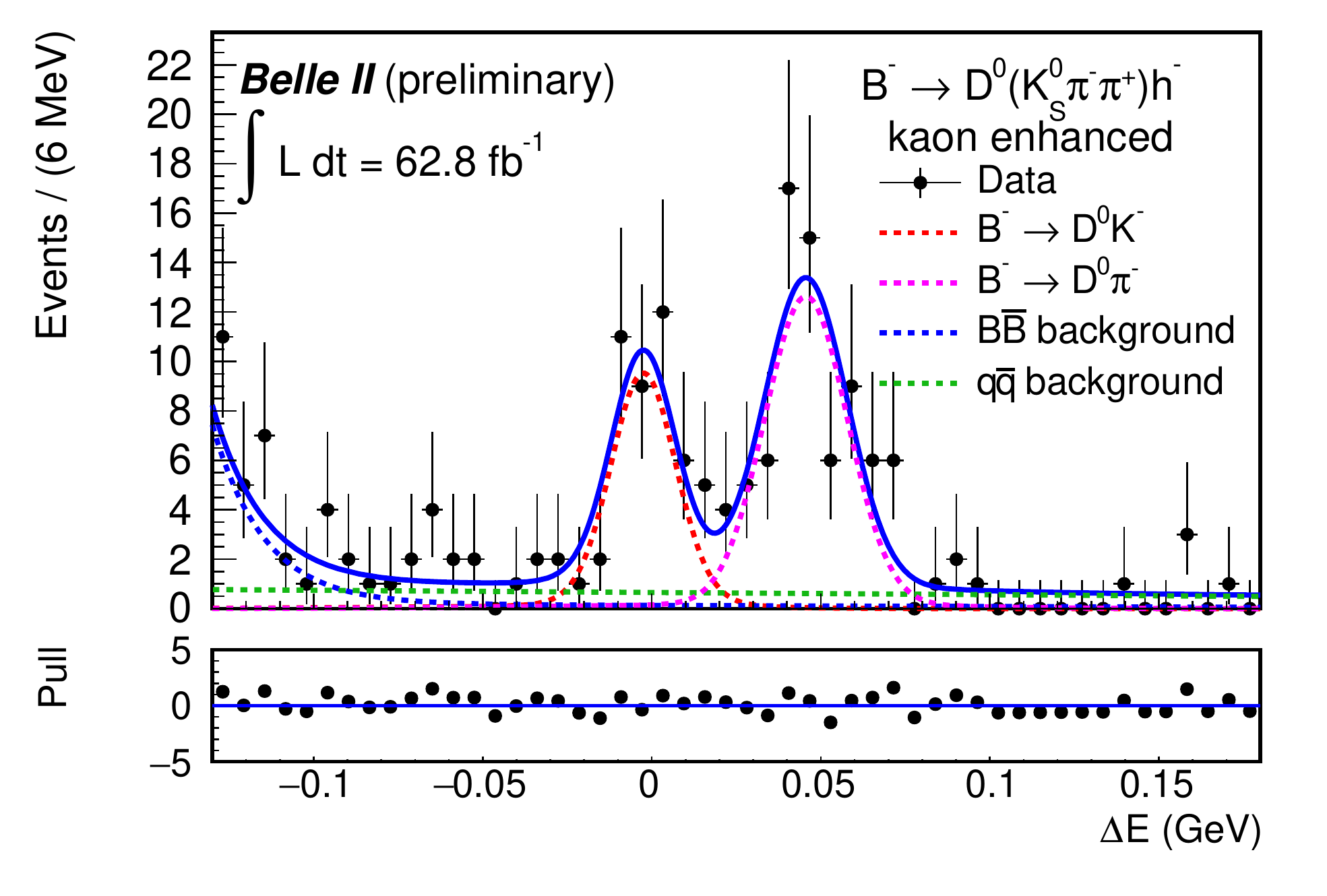} &
    \includegraphics[scale=0.4]{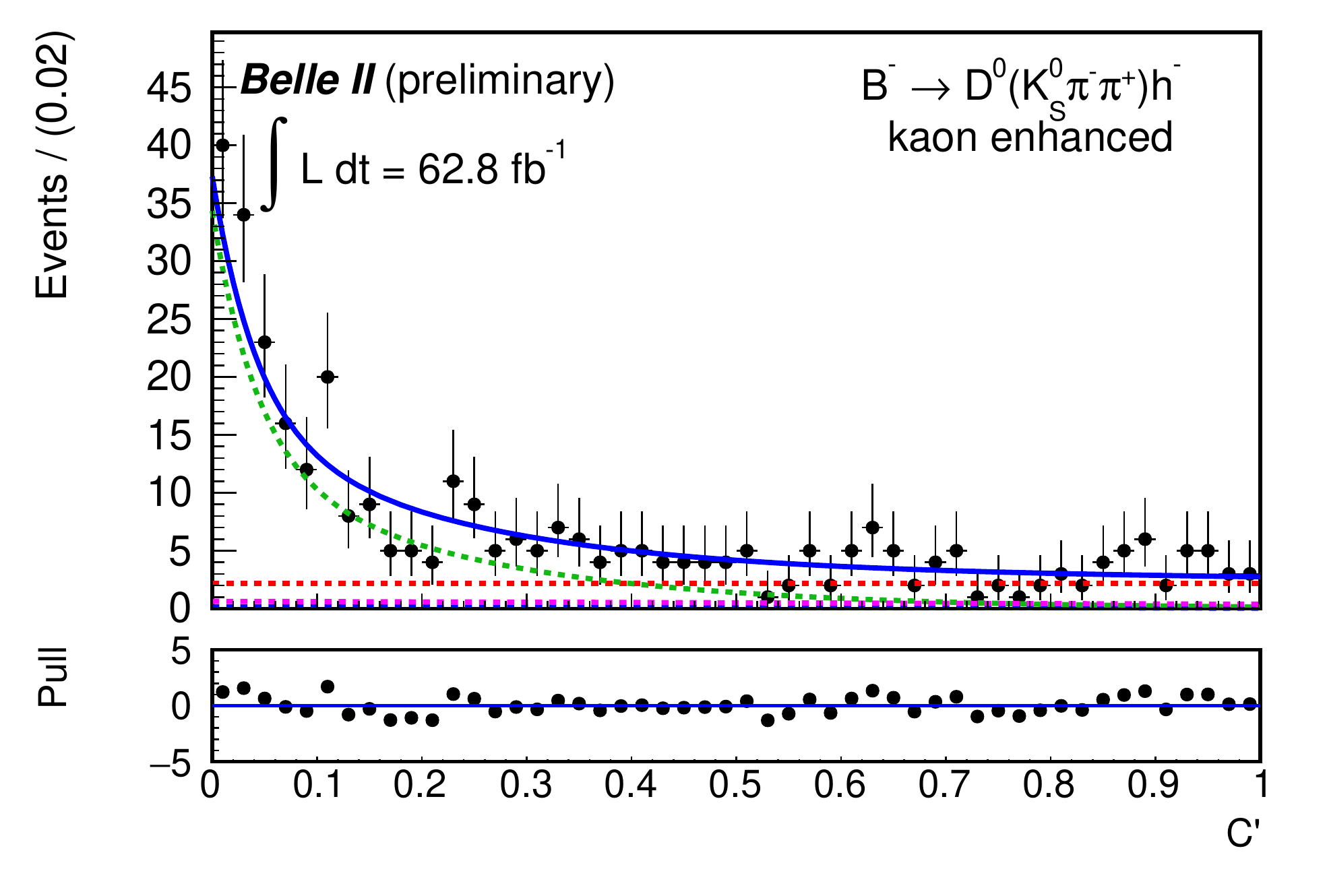} \\
\end{tabular}
\vspace{-0.2in}
 \caption{Signal-enhanced (left) $\Delta E$ and (right) $C'$ distributions for $B^- \to D^0(K_{\rm S}^0\pi^+\pi^-)h^-$ for the pion-enhanced (top) and kaon-enhanced (bottom) data sample. The projection of the total and individual components of a simultaneous unbinned  maximum-likelihood fit are overlaid. The signal-enhancement is achieved by requiring $\vert\Delta E\vert<0.03$~GeV and $0.65 < C' < 1$ on the $C^{\prime}$ and $\Delta E$ distributions, respectively.}
  \label{fig:Dh_Kspipi}
\end{figure}

Three background components are considered:
\begin{itemize}
    \item  continuum $q\bar{q}$ background;
    \item combinatorial $B\bar{B}$ background, in which the final state particles could be coming from both $B$ mesons in an event; and
    \item cross-feed peaking background from $B^+ \to Dh^+$, where~$h$~=~$\pi$,~$K$, in which the charged kaon is misidentified as a pion or vice versa.
\end{itemize}
There is no significant correlation between $\Delta E$ and $C'$, so the two-dimensional PDF for each of the components is the product of one-dimensional $\Delta E$ and $C'$ PDFs. 
The sum of a double Gaussian function and  an asymmetric Gaussian function with a common mean is used as the PDF to model the $\Delta E$ signal component in both samples. 
A uniform distribution is used to model the $C'$ signal component in both samples. 
The continuum background distribution is modeled with a first-order polynomial in $\Delta E$ and by the sum of two exponential  functions in $C'$. 
The $\Delta E$ distribution of combinatorial $B\bar{B}$ background in $D\pi$ is described by an exponential function.  
A first-order polynomial is added to the above two PDFs in the case of $B \to DK$ decays. 
The $C'$ distribution in the $B\to D\pi$ $(B\to DK)$ sample is modeled by a first-order (third-order) polynomial. The cross-feed peaking background in $\Delta E$ is modeled by the sum of a (double) Gaussian and  an asymmetric Gaussian in the $B\to DK$ $(B\to D\pi)$ sample and a first-order polynomial is used to model the $C'$ distribution for both samples.

All yields are determined from the fit to data.
For the $\Delta E$ PDFs the following parameters are determined in the fit to data: the signal PDF mean value, polynomial  coefficient for continuum background $\Delta E$ distribution, the exponential parameter for $B\bar{B}$ background, and the difference between the means of the signal and cross-feed peaks in the $B\to DK$ sample. For the $C'$ PDFs the following parameters are determined from the fit to data: the polynomial  coefficient of the $B\bar{B}$ background and one of the exponential parameters of the continuum background. All other shape parameters are fixed to those obtained from fits to appropriate MC samples. A scaling factor is applied on the $\Delta E$ signal resolution, which is a free parameter in the fit.
The signal-enhanced fit projections for the data are shown in Fig.~\ref{fig:Dh_Kspipi}, where the signal regions are defined as $\vert\Delta E\vert<~0.03$~GeV and $0.65 < C' < 1$.

\begin{figure}[!tb]
\begin{center}
\begin{tabular}{c c}
\includegraphics[width=8cm,height=6cm]{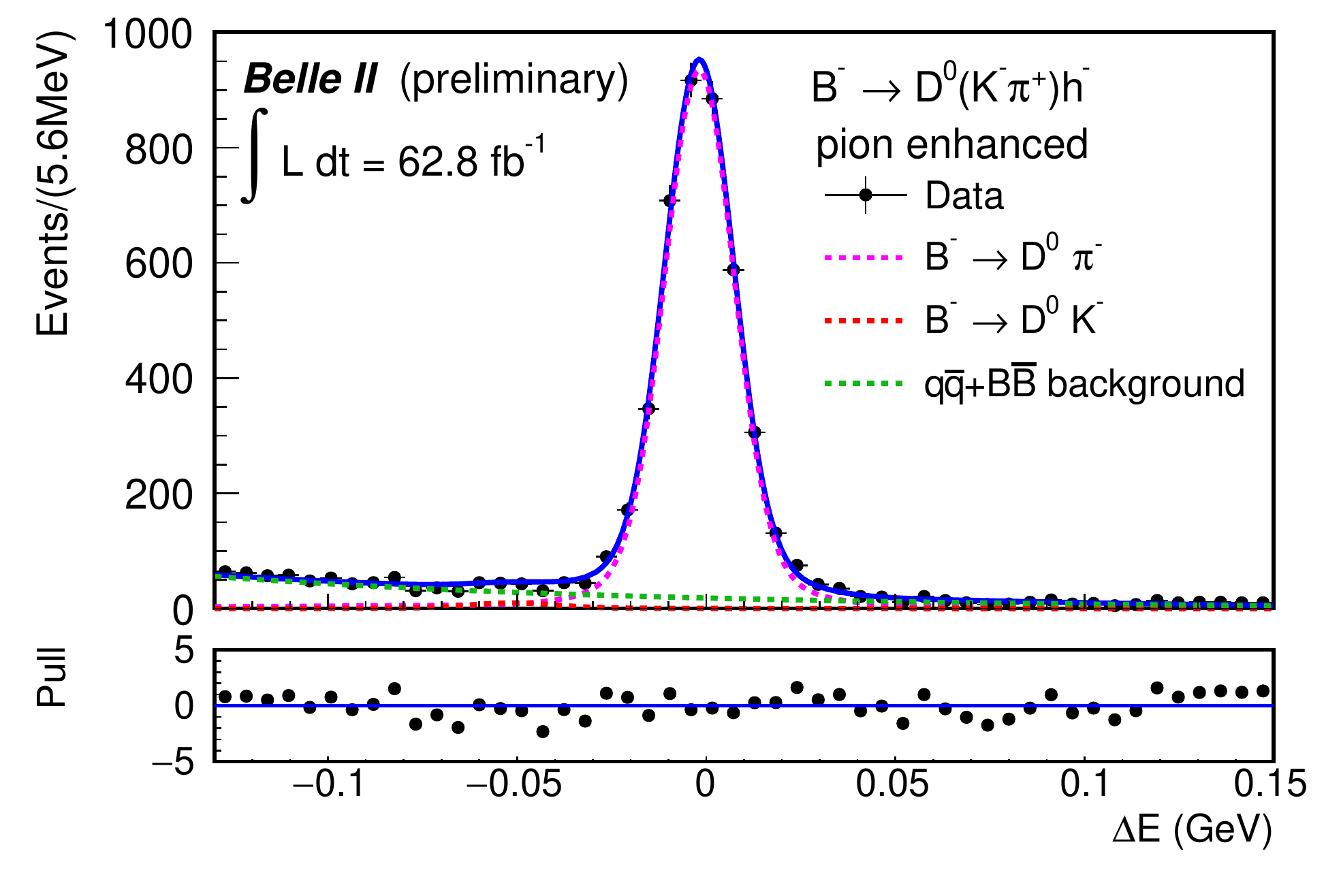} &
\includegraphics[width=8cm,height=6cm]{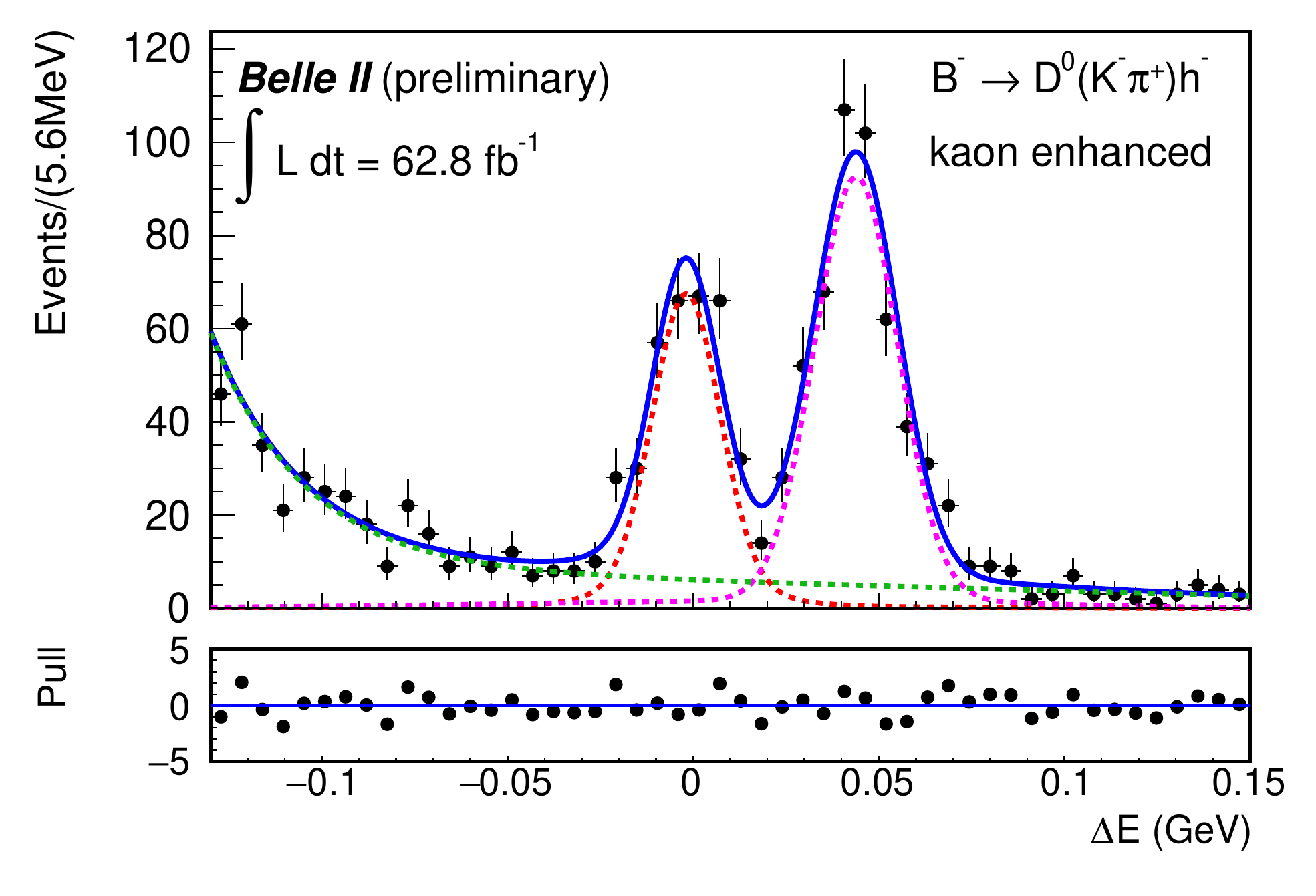} \\
\end{tabular}
\end{center}
  \caption{$\Delta E$ distributions for $B^- \to D^0(K^-\pi^+)h^-$ candidates that are (left) pion-enhanced and (right) kaon-enhanced from an 62.8~fb$^{-1}$ data sample. The projection of the total and individual components of a simultaneous unbinned  maximum-likelihood fit are overlaid.}
 \label{fig:Dh_kpi}
\end{figure}

\begin{figure}[!tb]
\begin{center}
\begin{tabular}{c c}
\includegraphics[width=8cm,height=6cm]{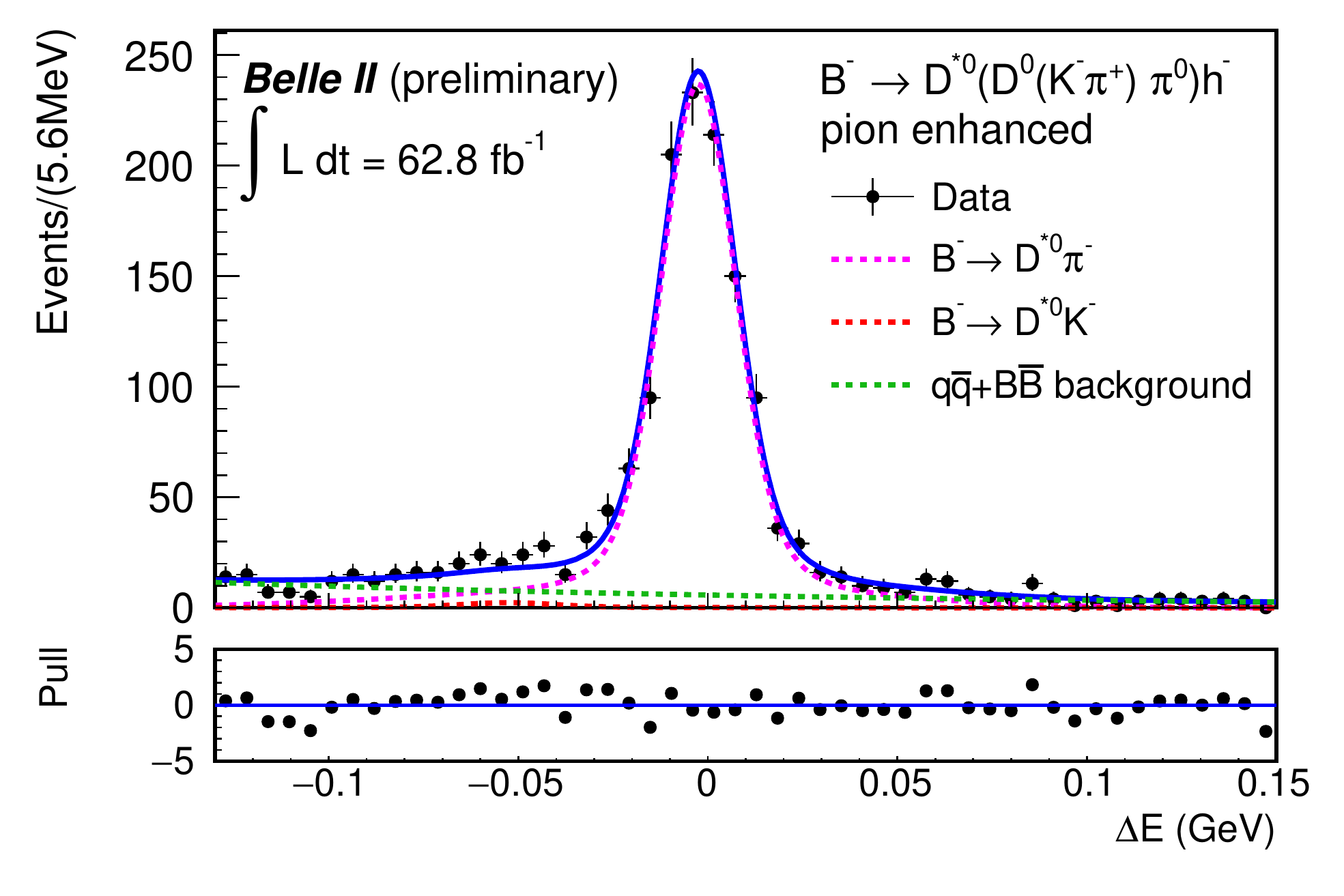} &
\includegraphics[width=8cm,height=6cm]{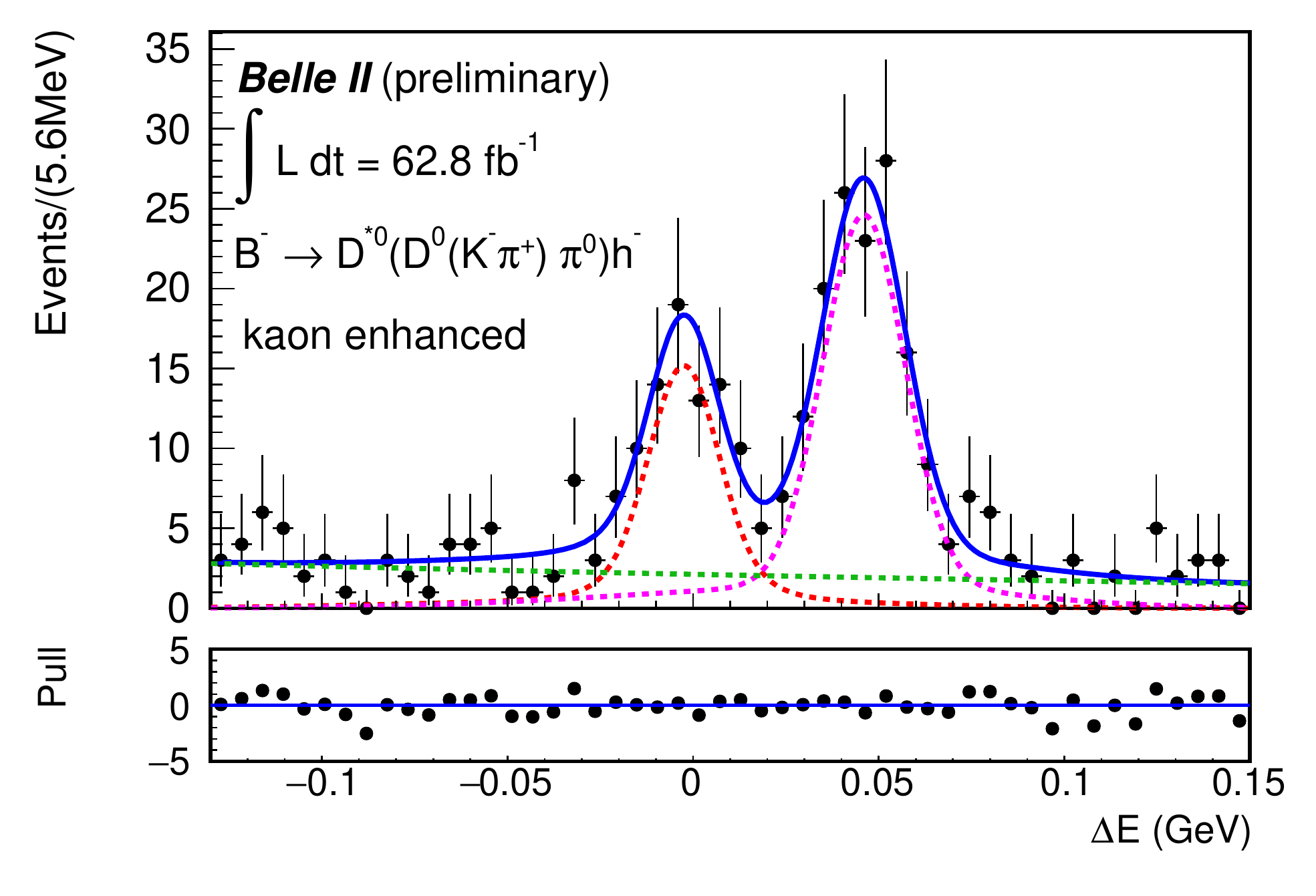} \\
\end{tabular}
\end{center}
  \caption{$\Delta E$ distributions for $B^- \to D^{*0}(D^0(K^-\pi^+)\pi^0)h^-$ candidates that are (left) pion-enhanced and (right) kaon-enhanced from an 62.8~fb$^{-1}$ data sample. The projection of the total and individual components of a simultaneous unbinned  maximum-likelihood fit are overlaid.}
 \label{fig:Dst0h_kpi}
\end{figure}

\begin{figure}[!tb]
    \centering
    \includegraphics[width=0.49\linewidth,page=1]{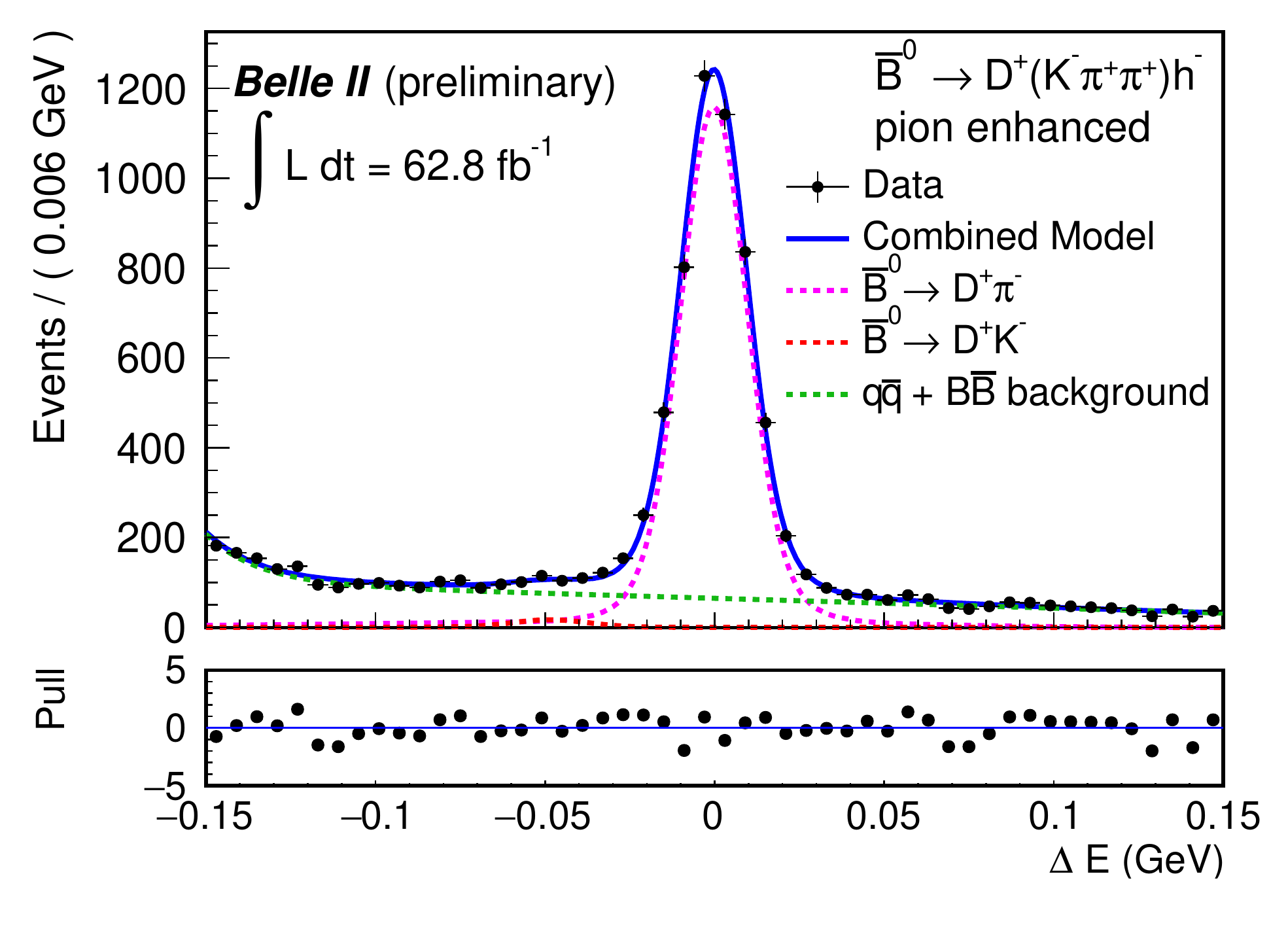}
    \includegraphics[width=0.49\linewidth,page=2]{simfit_winter2021_Kpipi.pdf}
\caption{$\Delta E$ distributions for $\bar{B}^0 \to D^+(K^-\pi^+\pi^+)h^-$ candidates that are (left) pion-enhanced and (right) kaon-enhanced from an 62.8~fb$^{-1}$ data sample. The projection of the total and individual components of a simultaneous unbinned  maximum-likelihood fit are overlaid.}
    \label{fig:Dh_kpipi}
\end{figure}

\begin{figure}[!tb]
\begin{center}
\begin{tabular}{c c}
\includegraphics[width=8cm]{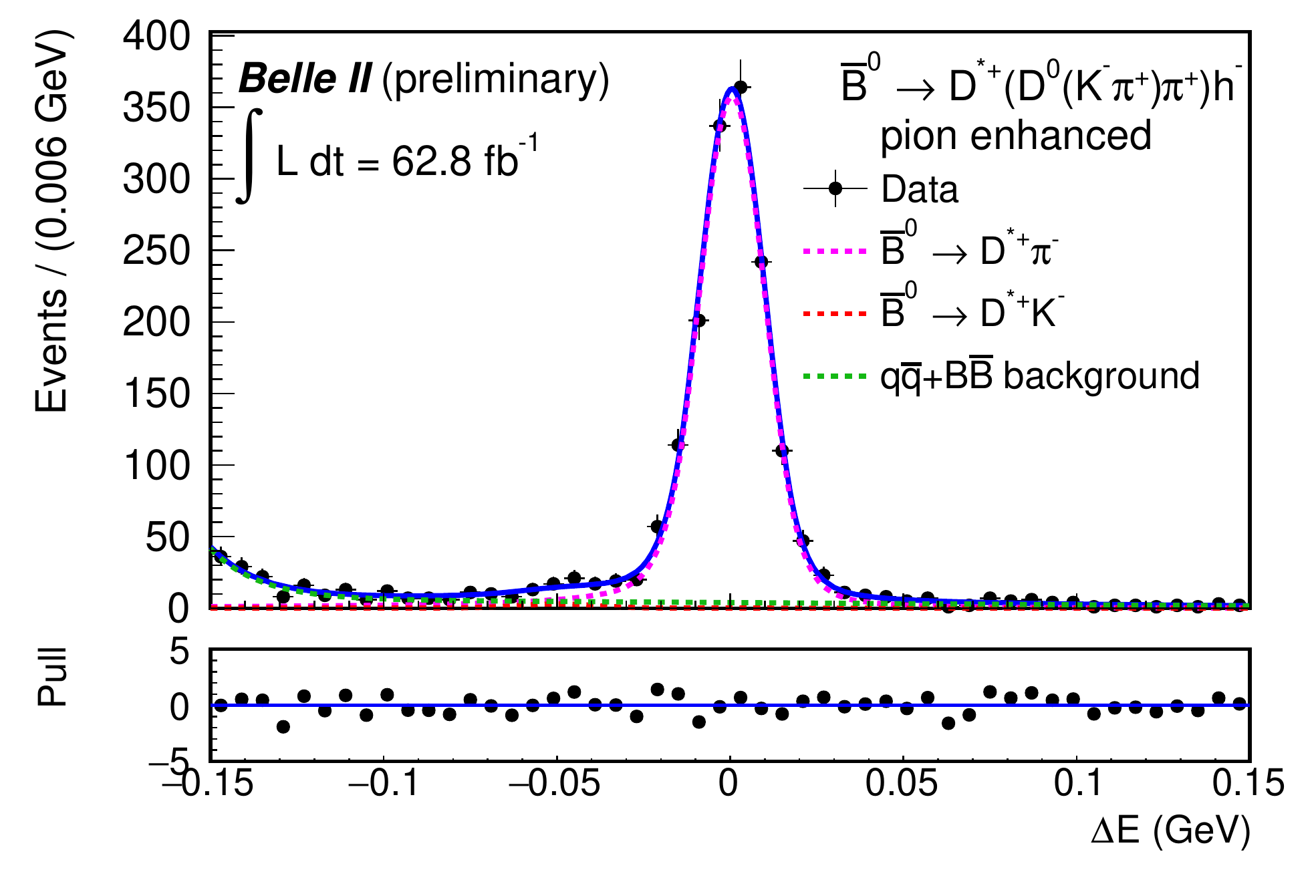} 
\includegraphics[width=8cm]{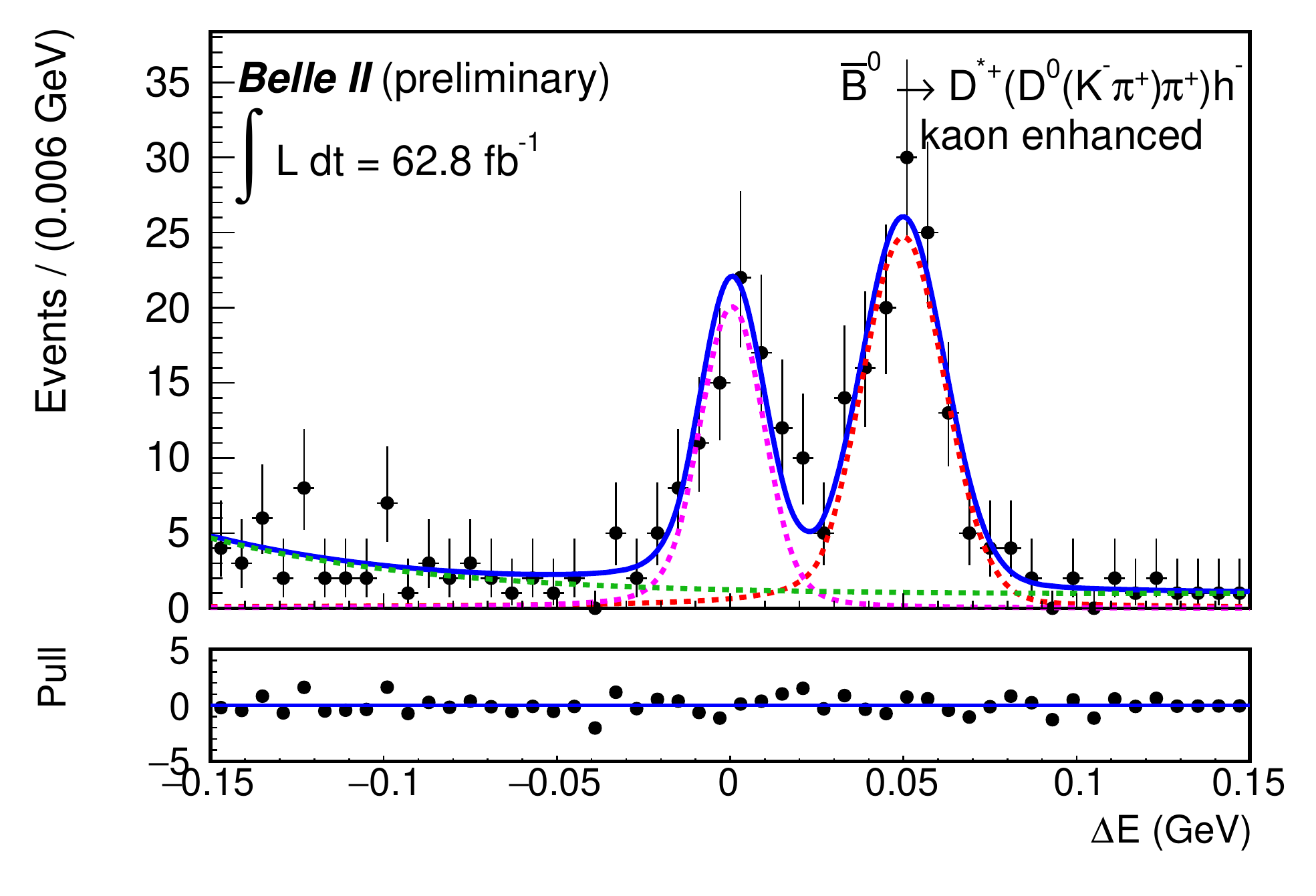}
\end{tabular}
\end{center}
 \caption{$\Delta E$ distributions for $\bar{B}^0 \to D^{*+}(D^0(K^-\pi^+)\pi^+)h^-$ candidates that are (left) pion-enhanced and (right) kaon-enhanced data sample from an 62.8~fb$^{-1}$ data sample. The projection of the total and individual components of a simultaneous unbinned  maximum-likelihood fit are overlaid.}
 \label{fig:Dsth_kpi}
\end{figure}

For  the other modes, in which continuum background is suppressed by  simply requiring $R_2 < 0.3$, the signal  yield is extracted using a simultaneous fit to only the $\Delta E$ distributions in both samples. In few modes,  there remains a peaking background from charmless  hadronic $B$ decays which  have the same final state  as the signal,  e.g. $B \to K \rho$ for the $B \to D (K\pi) \pi$ mode. The peaking background yield is fixed from MC simulation properly scaled by their measured branching  fraction~\cite{pdg}. The fit projections for the data are shown for $B^-\to D^{0}(K^-\pi^+)h^-$, $B^-\to D^{*0}[D^0(K^-\pi^+)\pi^0]h^-$, $\bar{B}^0\to D^{+}(K^-\pi^+\pi^+)h^-$ and $\bar{B}^0\to D^{(*)+}h^-$ in Figs.~\ref{fig:Dh_kpi},~\ref{fig:Dst0h_kpi},~\ref{fig:Dh_kpipi}~and~\ref{fig:Dsth_kpi}, respectively. The measured values of $R^{(*)+}$ and $R^{(*)0}$ are  listed in Tables~\ref{tab:R_value_final}~and~\ref{tab:R_value_final_D*}, respectively. The results reported by the LHCb Collaboration \cite{Aaij:2017ryw,Aaij:2013qqa,Aaij:2013xca} are also given in these tables; these measurements dominate the world-average values reported by the Particle Data Group (PDG)~\cite{pdg}.

%

%

\begin{table*}[!tb]
\centering
\renewcommand{\arraystretch}{1.3}
\caption{$R^{+}$ and $R^0$ results compared to those reported by  the LHCb Collaboration \cite{Aaij:2017ryw,Aaij:2013qqa}.}

\begin{tabular}{llll}
\hline\hline 
{} & ~~~$B^-\to D^0(K^-\pi^+)h^-$ & ~~~$\B^-\to D^0(K_{\rm S}^{0}\pi^+\pi^-)h^-$ & ~~~$\bar{B}^0\to D^{+}h^-$ \\ \hline
Belle II $R^{+/0}~(\times 10^{-2})$ & ~~$7.66 \pm 0.55$ ${}^{+0.11}_{-0.08}$ & ~~~$6.32 \pm 0.81$ ${}^{+0.09}_{-0.11}$ &$9.22 \pm 0.58 \pm 0.09$  \\
LHCb $R^{+/0}~(\times 10^{-2})$ & ~~$7.77 \pm 0.04 \pm 0.07$ \cite{Aaij:2017ryw} &  ~~~$7.77 \pm 0.04 \pm 0.07$ \cite{Aaij:2017ryw} & $8.22 \pm 0.11 \pm 0.25$ \cite{Aaij:2013qqa} \\
\hline\hline  
\end{tabular}
\label{tab:R_value_final}
\end{table*}

\begin{table*}[!tb]
\centering
\renewcommand{\arraystretch}{1.3}
\caption{$R^{*+}$ and $R^{*0}$ results compared to those reported by the  LHCb Collaboration \cite{Aaij:2017ryw,Aaij:2013xca}.}
\begin{tabular}{lll}
\hline\hline 
{} & ~~~~~~~$B^-\to D^{*0}h^-$  &  ~~~~~~$\bar{B}^0\to D^{*+}h^-$ \\ \hline
Belle II $R^{*+/0}~(\times 10^{-2})$ &  ~~$6.80 \pm 1.01 \pm 0.07$  &~~$5.99 \pm 0.82$ ${}^{+0.17}_{-0.08}$   \\
LHCb $R^{*+/0}~(\times 10^{-2})$ &  ~~$7.93 \pm 0.11 \pm 0.56$ \cite{Aaij:2017ryw} & ~~$7.76 \pm 0.34 \pm 0.26$ \cite{Aaij:2013xca}\\
\hline\hline  
\end{tabular}
\label{tab:R_value_final_D*}
\end{table*}

We consider several sources of systematic uncertainties. 
We assume the sources are independent, such that the total systematic uncertainty is the sum in quadrature of the contributions from individual sources. The individual contributions to the $R^{(*)+/0}$ systematic uncertainties, as well as the total systematic uncertainties, are reported in Table~\ref{tab:R_syst}. 
The major three sources of systematic uncertainty are the momentum scale factor applied in  the reconstruction to account for the incorrect magnetic field mapping, the fixed parameters in the PDF shape, and the kaon efficiency. 
The momentum of  charged particle tracks is corrected from the calibration results obtained with invariant  masses of well-known resonances. Those corrections are varied within their uncertainty.
Those due to the $\Delta E$ PDFs for the $DK$ signal, the $D\pi$ signal, and the $D\pi$  cross-feed are evaluated by varying the shape parameters by $\pm 1\sigma$, by replacing common parameter ($\Delta E$ mean of the $D\pi$ and $DK$ components) by a different parameter. The uncertainties due to the kaon identification efficiency are obtained by varying the assumed values by their uncertainties as obtained in data from the tagged $D$ control samples.
The uncertainty due to the peaking background is obtained by varying its yield by the uncertainty in its estimation. 

\begin{table*}[!tb]
\centering
\renewcommand{\arraystretch}{1.4}
\caption{Systematic uncertainties for $R^{*+/0}$ measurements.}
\begin{tabularx}{1.0\linewidth}{lccccc}
\hline\hline 
~~~~~~{}~~~~~~  & $D^0(K_{\rm S}^0\pi^+\pi^-)h^-$& $D^0(K^-\pi^+) h^-$&~~~$D^{*0} h^-$~~~ & ~~~$D^{+} h^-$~~~ &~~~ $D^{*+} h^-$~~~  \\ \hline
Kaon identification  $(\times 10^{-2})$ &  $^{+0.008}_{-0.008} $& $^{+0.010}_{-0.011} $ & $^{+0.020}_{-0.019} $ & $^{+0.023}_{-0.015} $ &${}^{+0.014}_{-0.013}$\\  
Momentum correction $(\times 10^{-2})$ & $^{+0.065}_{-0.100} $ & $ {}^{+0.109}_{-0.064} $ &${}^{+0.016}_{-0.018}$ & $^{+0.054}_{-0.040}$  & $^{+0.161}_{-0.030} $\\
PDF shape  $(\times 10^{-2})$ &$^{+0.066}_{-0.053}$ & $^{+0.030}_{-0.034}$ & $^{+0.066}_{-0.064}$ &  $^{+0.093}_{-0.079}$ & 
$^{+0.039}_{-0.036}$ \\ 
{Cross-feed fraction} $(\times 10^{-2})$ & $-$& $^{+0.004}_{-0.005} $& $^{+0.028}_{-0.025} $ & $^{+0.009}_{-0.003}$ & ${}^{+0.017}_{-0.014}$\\ 
{Common mean}  $(\times 10^{-2})$ & $+0.025$ & $-0.039 $& $+0.004$  &$- 0.003$ & $-0.067$\\
Peaking background $(\times 10^{-2})$ & $-$& $\pm 0.013 $ &$-$ & $^{+0.022}_{-0.015}$ & ${}^{+0.002}_{-0.002}$\\
\hline
Total $(\times 10^{-2})$ & $^{+0.093}_{-0.114} $ & ${}^{+0.114}_{-0.084}$ & ${}^{+0.071}_{-0.068}$ & ${}^{+0.112}_{-0.091}$ & ${}^{+0.166}_{-0.083}$\\
\hline\hline
\end{tabularx}
\label{tab:R_syst}
\end{table*}

\section{Summary}
\label{sec:summary}

We  have reported measurements of the decay rate ratio between $B\to D^{(*)}K^-$ and $B\to D^{(*)}\pi^-$. We use data collected by  the Belle II experiment in 2019 and 2020, corresponding to 
$62.8~\mathrm{fb}^{-1}$ of integrated luminosity, collected at  the $\Upsilon(4S)$ resonance. The measurements reported are for the decay modes $B^-\to D^0 h^-$, $B^{-}\to D^{*0}h^-$, $\bar{B}^{0}\to D^{+} h^{-}$ and $\bar{B}^{0}\to D^{*+} h^{-}$, where $h=\pi$ or $K$. The results are compatible with the world-average values reported by the PDG \cite{pdg}.

 \section*{ACKNOWLEDGEMENTS}
We thank the SuperKEKB group for the excellent operation of the
accelerator; the KEK cryogenics group for the efficient
operation of the solenoid; the KEK computer group for
on-site computing support; and the raw-data centers at
BNL, DESY, GridKa, IN2P3, and INFN for off-site computing support.
This work was supported by the following funding sources:
Science Committee of the Republic of Armenia Grant No. 20TTCG-1C010;
Australian Research Council and research grant Nos.
DP180102629, 
DP170102389, 
DP170102204, 
DP150103061, 
FT130100303, 
and
FT130100018; 
Austrian Federal Ministry of Education, Science and Research,
Austrian Science Fund No. P 31361-N36, and
Horizon 2020 ERC Starting Grant no. 947006 ``InterLeptons''; 
Natural Sciences and Engineering Research Council of Canada, Compute Canada and CANARIE;
Chinese Academy of Sciences and research grant No. QYZDJ-SSW-SLH011,
National Natural Science Foundation of China and research grant Nos.
11521505,
11575017,
11675166,
11761141009,
11705209,
and
11975076,
LiaoNing Revitalization Talents Program under contract No. XLYC1807135,
Shanghai Municipal Science and Technology Committee under contract No. 19ZR1403000,
Shanghai Pujiang Program under Grant No. 18PJ1401000,
and the CAS Center for Excellence in Particle Physics (CCEPP);
the Ministry of Education, Youth and Sports of the Czech Republic under Contract No.~LTT17020 and 
Charles University grants SVV 260448 and GAUK 404316;
European Research Council, 7th Framework PIEF-GA-2013-622527, 
Horizon 2020 ERC-Advanced Grants No. 267104 and 884719,
Horizon 2020 ERC-Consolidator Grant No. 819127,
Horizon 2020 Marie Sklodowska-Curie grant agreement No. 700525 `NIOBE,' 
and
Horizon 2020 Marie Sklodowska-Curie RISE project JENNIFER2 grant agreement No. 822070 (European grants);
L'Institut National de Physique Nucl\'{e}aire et de Physique des Particules (IN2P3) du CNRS (France);
BMBF, DFG, HGF, MPG, and AvH Foundation (Germany);
Department of Atomic Energy under Project Identification No. RTI 4002 and Department of Science and Technology (India);
Israel Science Foundation grant No. 2476/17,
United States-Israel Binational Science Foundation grant No. 2016113, and
Israel Ministry of Science grant No. 3-16543;
Istituto Nazionale di Fisica Nucleare and the research grants BELLE2;
Japan Society for the Promotion of Science,  Grant-in-Aid for Scientific Research grant Nos.
16H03968, 
16H03993, 
16H06492,
16K05323, 
17H01133, 
17H05405, 
18K03621, 
18H03710, 
18H05226,
19H00682, 
26220706,
and
26400255,
the National Institute of Informatics, and Science Information NETwork 5 (SINET5), 
and
the Ministry of Education, Culture, Sports, Science, and Technology (MEXT) of Japan;  
National Research Foundation (NRF) of Korea Grant Nos.
2016R1\-D1A1B\-01010135,
2016R1\-D1A1B\-02012900,
2018R1\-A2B\-3003643,
2018R1\-A6A1A\-06024970,
2018R1\-D1A1B\-07047294,
2019K1\-A3A7A\-09033840,
and
2019R1\-I1A3A\-01058933,
Radiation Science Research Institute,
Foreign Large-size Research Facility Application Supporting project,
the Global Science Experimental Data Hub Center of the Korea Institute of Science and Technology Information
and
KREONET/GLORIAD;
Universiti Malaya RU grant, Akademi Sains Malaysia and Ministry of Education Malaysia;
Frontiers of Science Program contracts
FOINS-296,
CB-221329,
CB-236394,
CB-254409,
and
CB-180023, and SEP-CINVESTAV research grant 237 (Mexico);
the Polish Ministry of Science and Higher Education and the National Science Center;
the Ministry of Science and Higher Education of the Russian Federation,
Agreement 14.W03.31.0026;
University of Tabuk research grants
S-0256-1438 and S-0280-1439 (Saudi Arabia);
Slovenian Research Agency and research grant Nos.
J1-9124
and
P1-0135; 
Agencia Estatal de Investigacion, Spain grant Nos.
FPA2014-55613-P
and
FPA2017-84445-P,
and
CIDEGENT/2018/020 of Generalitat Valenciana;
Ministry of Science and Technology and research grant Nos.
MOST106-2112-M-002-005-MY3
and
MOST107-2119-M-002-035-MY3, 
and the Ministry of Education (Taiwan);
Thailand Center of Excellence in Physics;
TUBITAK ULAKBIM (Turkey);
Ministry of Education and Science of Ukraine;
the US National Science Foundation and research grant Nos.
PHY-1807007 
and
PHY-1913789, 
and the US Department of Energy and research grant Nos.
DE-AC06-76RLO1830, 
DE-SC0007983, 
DE-SC0009824, 
DE-SC0009973, 
DE-SC0010073, 
DE-SC0010118, 
DE-SC0010504, 
DE-SC0011784, 
DE-SC0012704, 
DE-SC0021274; 
and
the Vietnam Academy of Science and Technology (VAST) under grant DL0000.05/21-23.

\end{document}